\documentclass{pasj00}
\draft

\begin{document}
\SetRunningHead{Author(s) in page-head}{Running Head}

\title{Long-term photometric behavior of the eclipsing cataclysmic variable V729 Sgr}



%
 \author{Han \textsc{Zhongtao},\altaffilmark{1,2,3}
         Qian  \textsc{Shengbang},\altaffilmark{1,2,3}
         Laj\'{u}s, E. \textsc{Fern\'{a}ndez},\altaffilmark{4,5}
         Voloshina \textsc{Irina},\altaffilmark{6}
         Zhu \textsc{Liying}\altaffilmark{1,2,3}}
\altaffiltext{1}{Yunnan Observatories, Chinese Academy of Sciences, PO Box 110, 650011 Kunming, China}
 \email{zhongtaohan@ynao.ac.cn}
\altaffiltext{2}{Key Laboratory of the Structure and Evolution of Celestial Objects, Chinese Academy of Sciences, PO Box 110, 650011 Kunming, China}
\altaffiltext{3}{Graduate University of the Chinese Academy of Sciences, Yuquan Road 19, Sijingshang Block, 100049 Beijing City, China}
\altaffiltext{4}{Facultad de Ciencias Astron\'{o}micas y Geof\'{i}sicas, Universidad Nacional de La Plata, Paseo del Bosque s/n, 1900, La Plata, Pcia. Bs. As., Argentina}
\altaffiltext{5}{Instituto de Astrof\'isica de La Plata (CCT La plata - CONICET/UNLP), Argentina}
\altaffiltext{6}{Sternberg Astronomical Institute, Moscow State University, Universitetskij prospect 13, Moscow 119992, Russia}
\KeyWords{binaries: eclipsing -- binaries : cataclysmic variables -- stars: individual (V729 Sgr)} 

\maketitle

\begin{abstract}
We present the analysis results of an eclipsing cataclysmic variable (CV) V729 Sgr, based on our observations and AAVSO data.
Some outburst parameters were determined such as outburst amplitude ($A_{n}$) and recurrence time ($T_{n}$), and then the relationship between $A_{n}$ and $T_{n}$ is discussed.
A cursory examination for the long-term light curves reveals that there are small-amplitude outbursts and dips present, which is similar to the behaviors seen in some nova-like CVs (NLs). More detailed inspection suggests that the outbursts in V729 Sgr may be Type A (outside-in) with a rise time $\sim1.76$ d. Further analysis also shows that V729 Sgr is an intermediate between dwarf nova and NLs, and we constrain its mass transfer rate to $1.59\times10^{-9}<\dot{M}_{2}<5.8\times10^{-9}M_{\odot}yr^{-1}$ by combining the theory for Z Cam type stars with observations. Moreover, the rapid oscillations in V729 Sgr were detected and analyzed for the first time. Our results indicate that the oscillation at $\sim 25.5$ s is a true DNO, being associated with the accretion events. The classification of the oscillations at $\sim 136$ and $154$ s as lpDNOs is based on the relation between $P_{lpDNOs}$ and $P_{DNOs}$. Meanwhile, the QPOs at the period of hundreds of seconds are also detected.
\end{abstract}

\section{Introduction}

Dwarf novae (DN) are a subclass of cataclysmic variables (CVs) which exhibit repetitive eruptions of amplitudes of $2-6$ mag, with some rare objects (e.g WZ Sge) with up to 8 mag range (Osaki 1996). Their outbursts are much smaller in amplitude and higher in frequency than the classical novae (CN).
V729 Sgr was identified as an eclipsing dwarf nova by Cieslinski et al. (2000). As early as 1928, H. van Gent (1932) discovered variability of the star. Follow up observations by Ferwerda (1934) shown that this object may be an irregular variable. It was first classified as a cataclysmic variable by using the photometric results in Cieslinski et al. (1997) and the spectroscopic results in Cieslinski et al. (1998). Cieslinski et al. (2000) further studied more details, publishing a lot of eclipsing light curves and deriving an orbital period of 0.1734055 d. In addition, these authors also estimated the outburst amplitude and recurrence time using their data together with published photometric data. Finally, V729 Sgr were classified as a possible Z Cam type dwarf nova. In fact, the outburst nature of this star are quite complex and little is known about it to date.

Apart from the large scale changes of the outburst itself in DN, there are rapid brightness variations of moderate coherence such as
dwarf nova oscillations (DNOs) with periods in the range from several seconds to tens of seconds and quasi-period oscillations (QPOs) with a larger amplitude and much longer period,
ranging from a few minutes to several thousand seconds. For the rapid oscillations in brightness of CVs, a series of papers (Woudt \& Warner 2002, Paper I; Warner \&
Woudt 2002, Paper II; Warner, Woudt \& Pretorius 2003, Paper III;
Warner \& Woudt 2006, Paper IV; Pretorius, Warner \& Woudt 2006,
Paper V; Warner \& Pretorius 2008, Paper VI; Woudt \& Warner 2009 Paper VII; Woudt et al. 2010,
Paper VIII) have been published. In general, the most of DN in outbursts and many nova-like variables (NLs) exhibit such variations. DNOs and QPOs in V729 Sgr therefore were analyzed for the first time by us in order to understand the accretion disc and accreted events.

In present paper, we will focus on two aspects of V729 Sgr's nature. First, combining the data from the American Association of Variable Star Observers (AAVSO) with our observations, the outburst properties of V729 Sgr were discussed. Second, rapid oscillations in the system were detected for the first time by using the Fourier transform method in the modifications by Lomb (1976) and Scargle (1982).

\section{Observations and Data Preparation}

V729 Sgr was monitored photometrically from 2010 to 2015 by using the 2.15-m Jorge Sahade telescope (JST) mounted a Roper Scientific, Versarray 1300B camera with a thinned EEV CCD36-40 de $1340\times1300$ pix CCD chip at Complejo Astron\'omico E1 Leoncito (CASLEO), San Juan, Argentina. During the observations, $I-$band was applied on 07, June 2010 and 05, April 2011, and no filters were used for other data. All observed CCD images were performed by using the aperture photometry package of IRAF. Integration time is $\sim5-15$ s. Differential photometry was performed, with a nearby non-variable comparison star. For all observations, the same comparison star was used to calculate the relative brightness.
All of observations during 2010-2015 are shown in Fig. 1. The light curves were phased by following ephemeris,
\begin{equation}
Min.HJD=2456889.6965+0.1734055\times{E}, \label{linear ephemeris}
\end{equation}
where HJD2456889.6965 is the initial epoch from our mid-eclipse times observed on 8 August 2014, 0.1734055 d is the orbital period provided by Cieslinski et al. (2000). Our observations exhibit at least three outburst episodes in Fig. 1. It is shown that there is an outburst with the amplitude of $\sim1.82$ mag in the time interval between July 20 and August 23, 2014.
To investigate the outburst properties, the long-term light curves were required. Fortunately, the AAVSO data of $\sim15$ yr provide a good opportunity to study its outbursts and other variations. The upper panel of Fig. 2 shows the full AAVSO light curves of V729 Sgr from 2000 July to 2015 June. More detailed analysis was given in Section 3.1.

\begin{figure}[!h]
\begin{center}
\includegraphics[width=16cm]{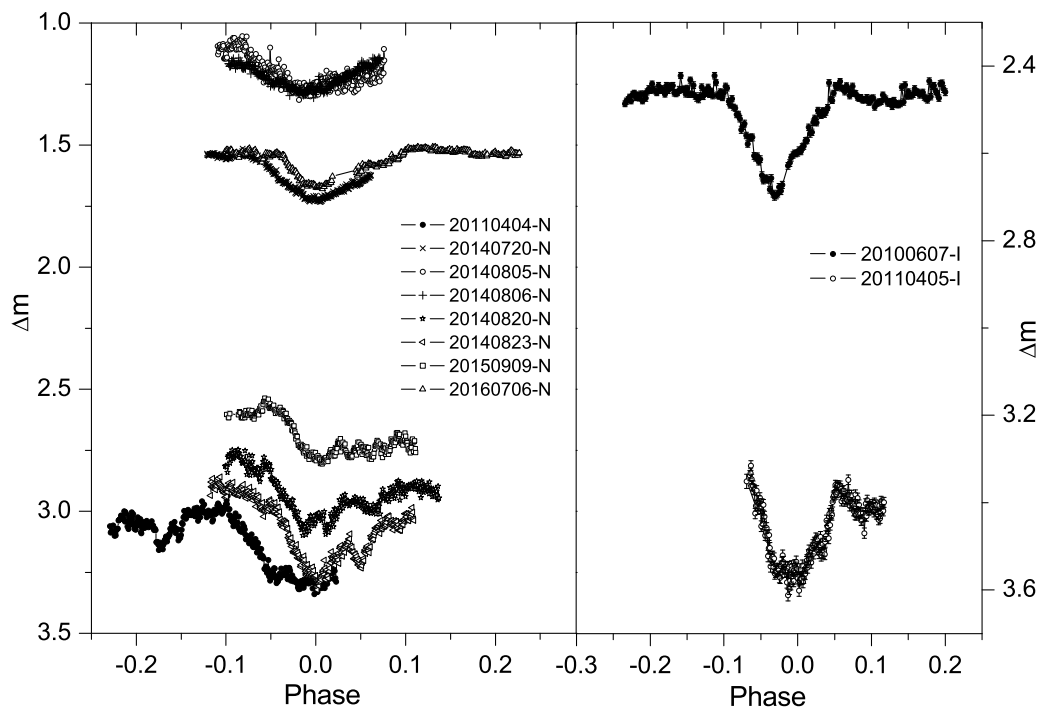}
\caption{All light curves of V729 Sgr plotted in phase using the equation (1). The light curves shown in the left panel were observed in N-band, the light curves shown in the right panel were obtained in I-band, respectively. The data exhibit at last three outburst episodes.}
\end{center}
\end{figure}

\section{Results and discussion}

\subsection{outburst properties}
The outburst amplitudes of $\sim1.0-1.5$ mag and recurrence time of several tens of days in V729 Sgr were reported by Cieslinski et al. (2000) using the published historical data.
However, the less historical data coverage may reduce the reliability of the conclusions.  Fortunately, all $783$ observations from AAVSO were offered to study its outburst (see Fig. 2). A detailed inspection for the upper panel displays the outbursts with amplitudes of $\sim0.6-2.2$ mag and the recurrence times of $\sim22-34$ d for long outburst and $\sim8.6-17$ d for short outburst.
The Lomb$-$Scargle power spectrum derived using the method of Lomb (1976) and Scargle (1982) for all data shows one strongest peak near $26.5$ d and some secondary peaks corresponding to the periods between $9.1$ and $15.6$ d. This implies that these prominent peaks may represent the typical recurrence times.
Kukarkin \& Parenago (1934) first noted that there was a relation between the outburst amplitude $A_n$ and the outburst recurrence time $T_n$ for DN and recurrence nova (RN). After that, this relation has done many times revised and improved to constrain its application range. For example, a general correlation was found by analysing DN normal outbursts from van Paradijs (1985). Moreover, this relation is also extended by including the orbital period (Richter \& Braeuer 1989). The most common version of K-P relation from Warner (1995) is as follows:
\begin{equation}
A_n=0.7(\pm0.43)+1.9(\pm0.22)\log T_n.
\end{equation}
This is an empirical relation. To explore if this relation is model dependent, moreover,  a theoretical K-P relation was derived by Kotko \& Lasota (2012) using the disc instability model (DIM):
\begin{equation}
A_n=C_1+2.5\log T_n,
\end{equation}
where the constant term $C_1=2.5\log 2\tilde{g}-2.5\log t_{dec}+BC_{max}-BC_{min}$, which depends on the properties of primary star and the viscosity parameter $\alpha$.
The parameters of outbursts in V729 Sgr and the K-P relations are plotted in Fig. 3. The open circles represent the statistical data from AAVSO database, the red solid line refer to empirical K-P relation and red dashed lines denote its upper and lower uncertainties. Note that $\log T_n$ in Fig. 3 have a larger range, which is consistent with the peaks near 26.5 d and between 9.1 and 15.6 d.
However, the observational data were not covered in this K-P relation. Note that this relation for only those systems with $A_{n}>2.5$ mag is significant (Warner 1995). For comparison, a least-squares linear fit to the data gives the K-P relation for V729 Sgr (blue solid line in Fig. 3):
\begin{equation}
A_n=-1.59(\pm0.48)+2.48(\pm0.29)\log T_n,
\end{equation}
This equation can also be expressed as:
\begin{equation}
A_n=1.65(\pm0.30)+2.48(\pm0.29)[\log T_n-1.31(\pm0.08)],
\end{equation}
where $1.65(\pm0.30)$ and $1.31(\pm0.08)$ are the mean values of $A_n$ and $\log{T_n}$, respectively.
The slope is larger than 1.9 of the empirical relation and is equal approximately to 2.5 of the theoretical relation. Thus, the theoretical K-P relation follows the overall trend of observational data in V729 Sgr reasonably well. The intercept ($C_1=-1.59$) should be connected with the nature of the system (see Kotko \& Lasota 2012). Finally, we suggest that the K-P relation may be model dependent and represents general properties of DN outburst.

\begin{figure}[!h]
\begin{center}
\includegraphics[width=18cm]{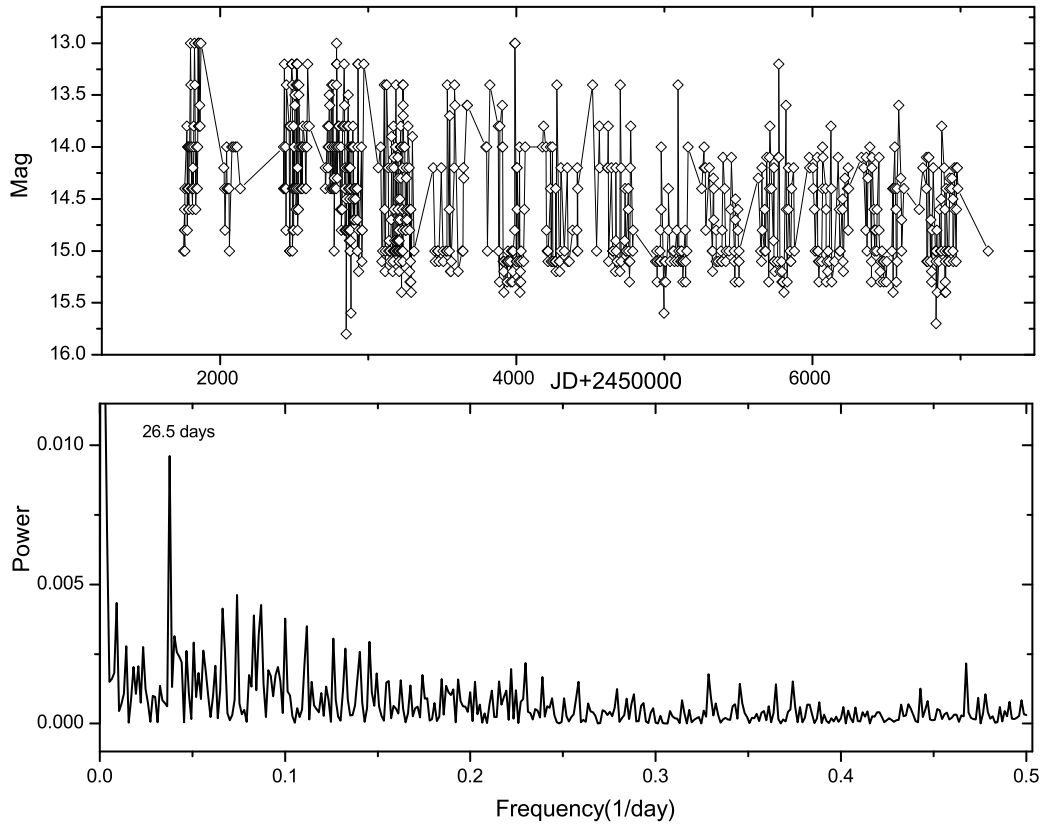}
\caption{Top panel: the long-term observations from AAVSO on V729 Sgr. Lower panel: the power spectrum derived using all data, the strongest peak corresponds to 26.5 days.}
\end{center}
\end{figure}

\begin{figure}[!h]
\begin{center}
\includegraphics[width=18cm]{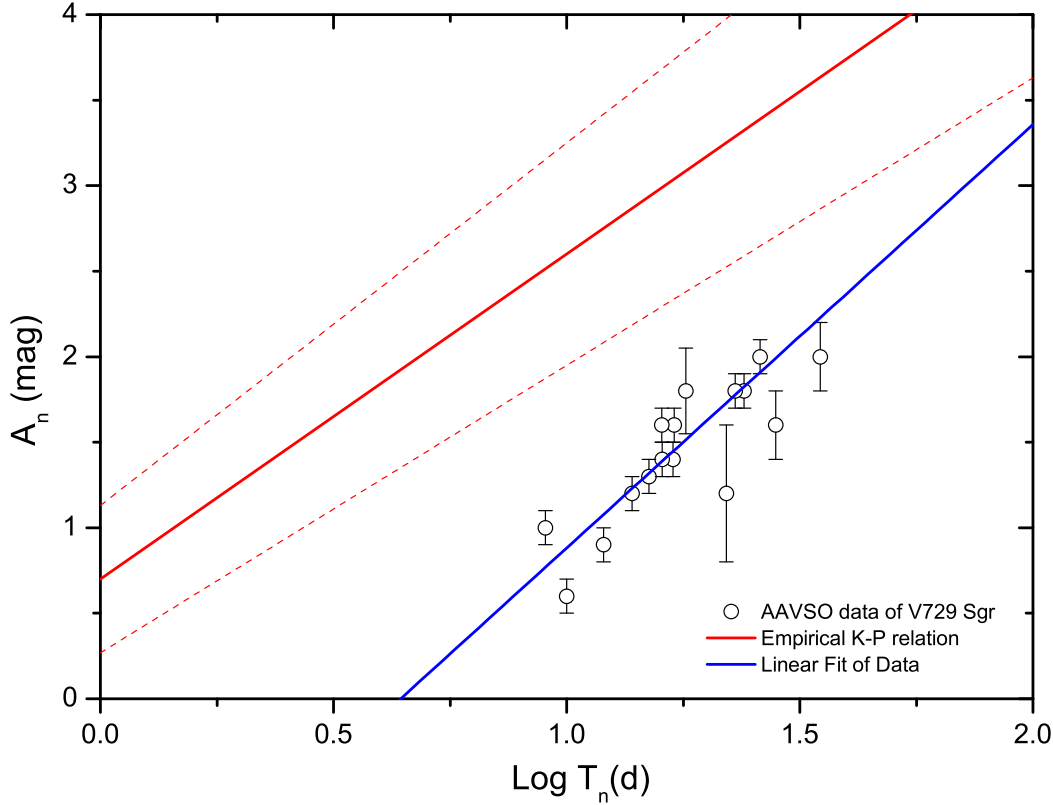}
\caption{Comparison of the K-P relation and AAVSO observational data for V729 Sgr. The solid circles represent the statistical data, the lines (dash and solid line) refer to upper and lower uncertainties of K-P relation. }
\end{center}
\end{figure}

\begin{figure}[!h]
\begin{center}
\includegraphics[width=18cm]{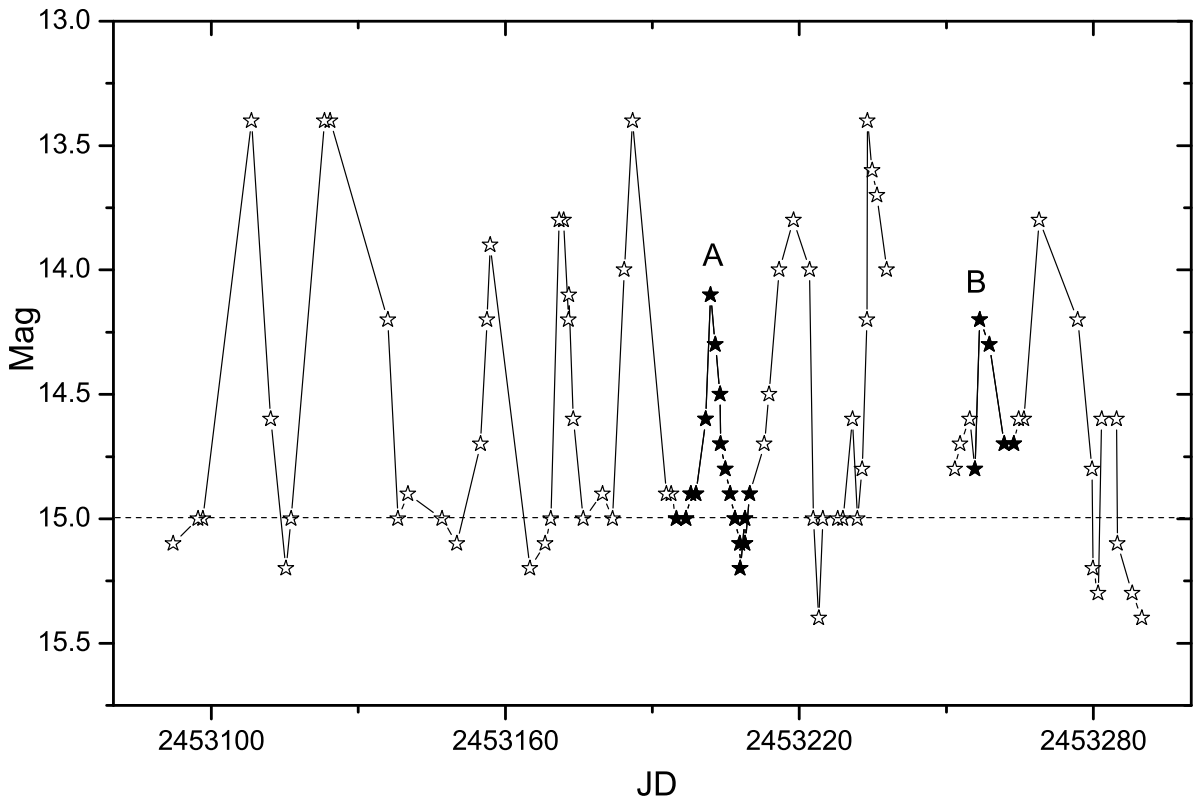}
\caption{Part of the light curve extracted from the upper panel in Fig. 2 during the period 2004 March to October. Apart from several normal outbursts, there are the smaller-amplitude outbursts (the solid asterisk) and dips.}
\end{center}
\end{figure}

\begin{figure}[!h]
\begin{center}
\includegraphics[width=20cm]{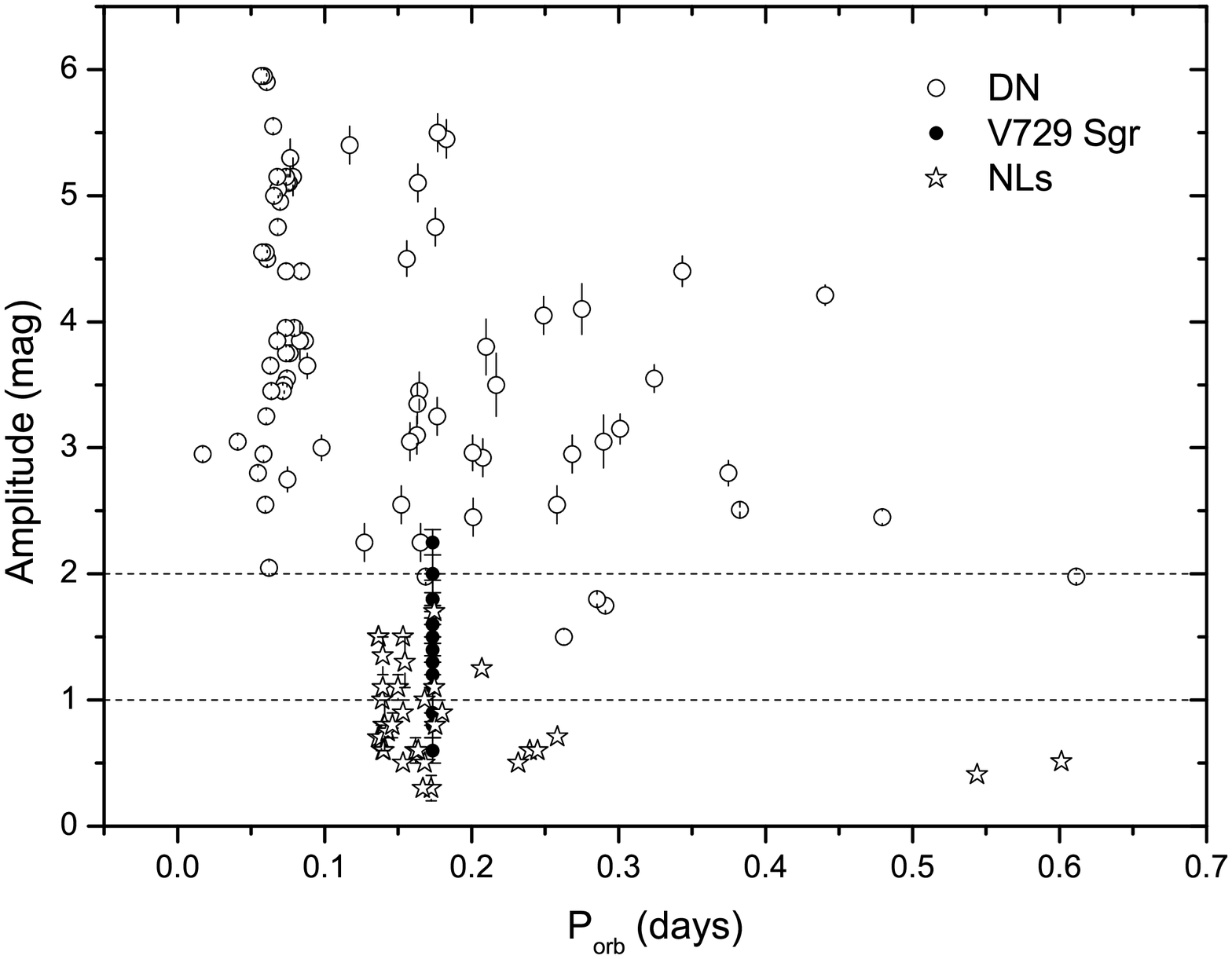}
\caption{A distribution of the amplitude of outburst or modulation in brightness in CVs using the data of both 77 DN and 33 NLs obtained from AAVSO observations and literatures. The position of V729 Sgr shown that it has a wide amplitude range and is thought to a link between DN and NLs.}
\end{center}
\end{figure}

Fig. 4 shows the partial data extracted from the upper panel in Fig. 2 during the period from 2004 March to October. Apart from several normal outbursts, we note the presence of two smaller-amplitude outbursts (the solid asterisk in the Fig. 4, labelled A and B.) with amplitudes of $\sim0.6$ and $0.9$ mag. The smaller-amplitude outbursts are similar to a few NLs which show "stunted" outbursts with amplitudes up to $1$ mag (e.g. Honeycutt et al. 1995, 1998; Warner 1995; Ramsay et al. 2016).
Another feature in Fig. 4 is that there are dips following the outbursts, i.e. the dips are paired with outbursts. Although the fragmentary data and uncertainty in the light curves, there is nevertheless an obvious dip accompanied outburst A (see Fig. 4). This dip has a depth of $\sim0.2$ mag with duration of 2 d. Honeycutt et al. (1998) reported that some dips exist in several old novae and NLs, which have a depth of $0.2-0.5$ mag with a FWHM ranging from 2 to 50 d. In fact, these dips have some resemblance to dips of a subtype of NLs VY Scl stars. Most recently, a NL KIC9202990 also shows two dips of depths $0.4$ and $0.6$ mag with FWHM of $2-3$ d (Ramsay et al. 2016). It is thought that these behaviors are caused by temporary reduction in mass transfer rate due to the activity of secondary star (e.g. Howell et al. 2000 and Kafka \& Honeycutt 2005).

More detailed examination for the AAVSO data reveals that there are diverse outburst behaviors and the rise speed are generally faster than the decay. We also used several more complete outbursts to estimate the parameters, which are listed in Table 1. It seems that the decay time is dependent on the magnitude at the maximum (see Table 1), which is similar the magnetic dwarf nova DO Dra (Andronov et al. 2008). The nature of a rapid rise and a slower decay indicated that the outbursts are outside-in outbursts (Type A) (Smak 1984). The rise time for Type A outburst is
\begin{equation}
t_{out-in}\sim\frac{R_{d}}{\alpha_{h}c_{s}}
\end{equation}
where $R_{d}$ is the radius of accretion disc, $\alpha_{h}c_{s}$ is the velocity of the inward-moving heating front and $c_{s}$ represents sound speed. The outer disc radii during outbursts in dwarf novae can be approximated to $R_{d}\sim0.9R_{L}$ (e.g. Smak 2001), where $R_{L}$ is the effective radius of the Roche Lobe of primary star derived by Eggleton (1983) as:
\begin{equation}
\frac{R_{L}}{a}=\frac{0.49q^{2/3}}{0.6q^{2/3}+\ln({1+q^{1/3}})}
\end{equation}
where $a$ is the orbital separation, $q=M_{1}/M_{2}$. Combining the physical parameters of V729 Sgr (Cieslinski et al. 2000) with Kepler's third law
\begin{equation}
{P^{2}_{orb}}=\frac{4\pi^{2}a^{3}}{G(M_{1}+M_{2})},
\end{equation}
to yield $a\approx9.6\times10^{10}$ cm. So, $R_{L}\approx0.44a=4.2\times10^{10}$ cm and $R_{d}\sim0.9R_{L}\approx3.8\times10^{10}$ cm. For typical parameters in CVs, $c_{s}\sim2.5\times10^6$ cm/s and $\alpha_{h}\sim0.1$ during outbursts (Frank et al. 2002). Applying these parameters and equation (5) we find $t_{out-in}\sim1.76$ d, in rough agreement with observations.

\begin{table*}
\caption{Parameters of rise and decay of several outbursts.}
\begin{center}
 \large
   \begin{tabular}{ccccc}\hline\hline
JD         \qquad \qquad &$V_{max}$(mag)    \qquad  \qquad  &$\tau_{rise}$(d/mag) \qquad \qquad   &$\tau_{decay}$(d/mag)  \qquad \qquad   &Type  \\\hline
2453170   \qquad \qquad  &13.6(0.3) \qquad  \qquad     &1.42(0.05)           \qquad \qquad   &3.33(0.11)             \qquad \qquad                     &A\\
2452787   \qquad \qquad  &13.4(0.3) \qquad  \qquad     &1.43(0.02)           \qquad \qquad   &2.86(0.07)             \qquad \qquad                    &A\\
2453550   \qquad \qquad  &13.7(0.3) \qquad  \qquad  &1.57(0.03) \qquad \qquad   &4.01(0.06) \qquad \qquad   &A\\
2454702   \qquad \qquad  &13.2(0.2) \qquad  \qquad     &1.88(0.06)           \qquad \qquad   &2.14(0.10)             \qquad \qquad                    &A\\
2452930   \qquad \qquad  &13.4(0.4) \qquad  \qquad     &2.22(0.08)           \qquad \qquad   &2.94(0.06)             \qquad \qquad                    &A\\
2453186   \qquad \qquad  &13.8(0.3) \qquad  \qquad     &2.58(0.02)           \qquad \qquad   &4.55(0.10)             \qquad \qquad                    &A\\
2453202   \qquad \qquad  &14.1(0.2) \qquad  \qquad     &3.65(0.05)           \qquad \qquad   &5.45(0.10)             \qquad \qquad                   &A\\\hline
\end{tabular}
\end{center}
\end{table*}

V729 Sgr was classified as a Z Cam type DN by Cieslinski et al. (2000). The Z Cam type stars generally have relatively high mass
transfer and accretion rates, and outburst amplitudes are lower than the most of dwarf novae (Szkody et al. 2013). A distribution of both the outburst and modulation amplitude in brightness of CVs is displayed in Fig. 5 using the data of both 77 DN obtained from AAVSO database and 33 NLs given by AAVSO and literatures (e.g. Honecutt et al. 1998, Honecutt et al. 2001, Gies et al. 2013, Honeycutt et al. 2014 and Ramsay et al. 2016 etc.), and the values of V729 Sgr are also added. It is shown that V729 Sgr may be an intermediate between DN and NLs. Many authors (e.g. Smak 1984, Cannizzo 1993, Osaki 1996, Lasota 2001 and Buat-M{\'e}nard et al. 2001) have described the model of Z Cam type stars' outbursts and standstill. The model pointed out that the systems with mass transfer rates below the critical value ($\dot{M}_{crit}$) generate DN outbursts, while those above the critical rate will become NLs. The Z Cam systems are thought to be fall on the boundary between the thermally stable NLs and thermally unstable DN. This may explain why V729 Sgr exhibits rich and complex behaviors such as the "stunted" outbursts and dips in NLs and the DN normal outbursts. By taking into account the importance of the critical mass transfer rate, it is very necessary to estimate it. Frank et al. (2002) give an expression for $\dot{M}_{crit}$:
\begin{equation}
\dot{M}_{crit}\simeq3\times10^{-9}(P_{orb}/{3})^2 M_{\odot}yr^{-1}
\end{equation}
where $P_{orb}$ is in hours. For V729 Sgr, we find $\dot{M}_{crit}\simeq5.8\times10^{-9}M_{\odot}yr^{-1}$. As noted above, the most of normal outbursts in V729 Sgr are Type A (outside-in). Osaki (1996) pointed out that when $1.59\times10^{-9}<\dot{M}_{2}<4.76\times10^{-9}M_{\odot}yr^{-1}$, the outburst is Type A; when $4.76\times10^{-9}<\dot{M}_{2}<1.59\times10^{-8}M_{\odot}yr^{-1}$, the recurrence time will be further shorten, this may corresponds to the "stunted" outbursts ($<1$ mag) in V729 Sgr. If $\dot{M}_{2}>5.8\times10^{-9}M_{\odot}yr^{-1}$ in V729 Sgr, however, it will across the instability boundary becoming NLs that remain in a constant high state, viz standstill. Unfortunately, so far, no sign of standstill was found. Therefore, $\dot{M}_{2}$ was restricted to the range from $1.59\times10^{-9}$ to $5.8\times10^{-9}M_{\odot}yr^{-1}$.

\subsection{DNOs and QPOs in V729 Sgr}
Except for the large-scale outburst brightness variations, there are rapid brightness oscillations of moderate coherence and low amplitude in CVs. They are generally divided into two types: DNOs and QPOs, which were most detected in the DN during outburst and many NLs. A review article have been given by Warner (2004). Normally, DNOs have low-amplitude oscillations in brightness with short periods, and QPOs are longer timescale modulations with larger amplitude. The relation between $P_{QPOs}$ and $P_{DNOs}$ is $P_{QPOs}\approx16\times P_{DNOs}$. Besides, Warner et al. (2003) also reported that there exist the longer-period DNOs (lpDNOs) with the typical intermediate periods between $P_{QPOs}$ and $P_{DNOs}$, and their relation is $P_{lpDNOs}\approx4\times P_{DNOs}$.

\begin{figure*}
\begin{center}
\includegraphics[width=14cm]{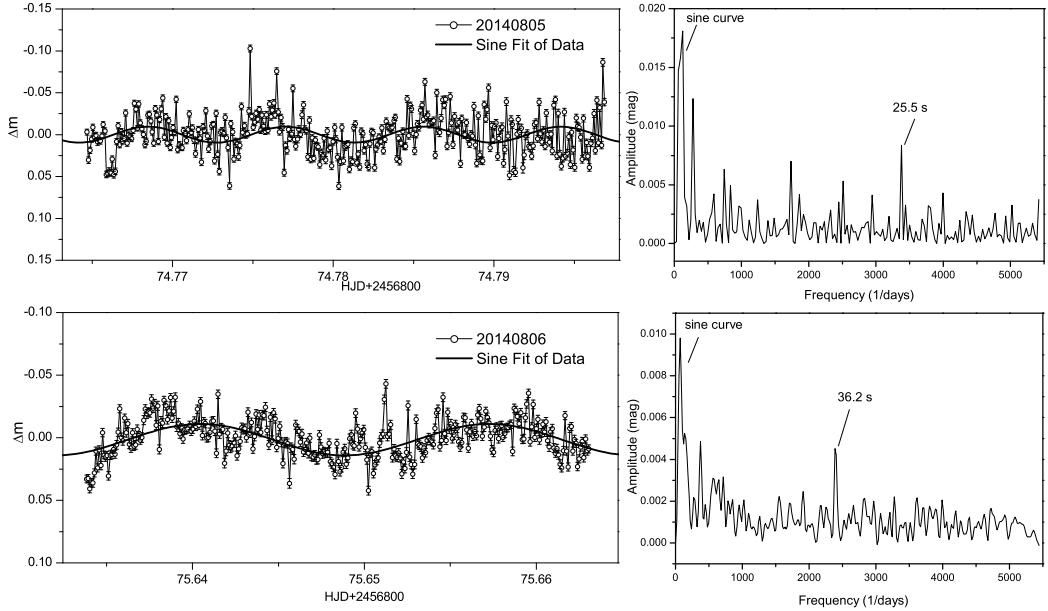}
\caption{The light curves and their Fourier spectrums near the outburst maximum on 5 and 6 August 2014.  }
\end{center}
\end{figure*}

\begin{figure*}
\begin{center}
\includegraphics[width=16cm]{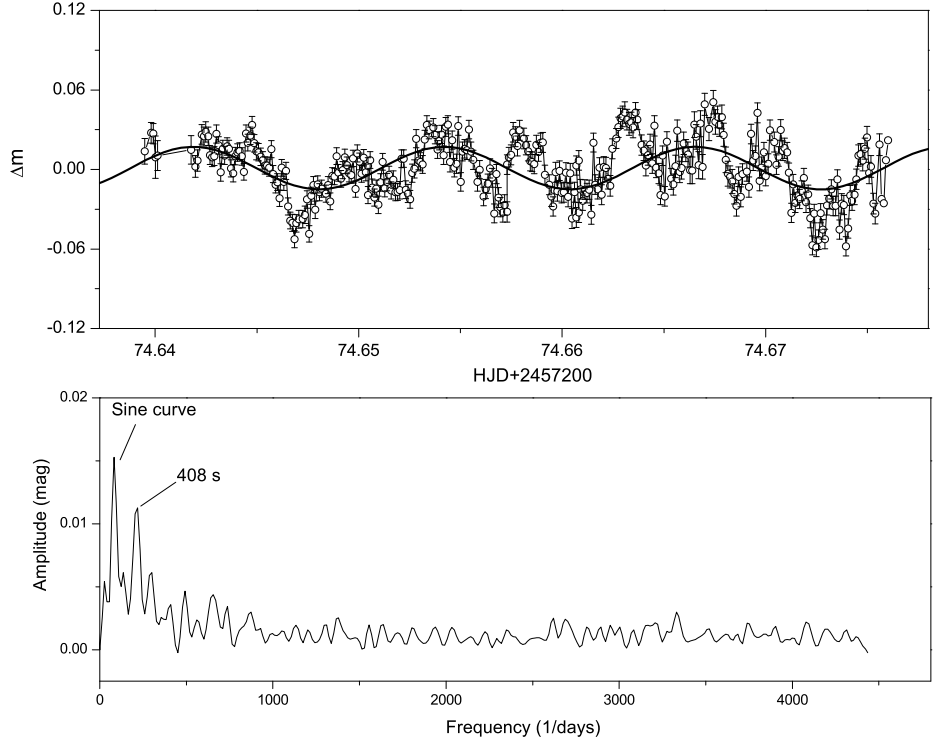}
\caption{The light curve and its Fourier spectrum on 9 September 2015. A QPO oscillation at $\sim408$ s is clearly visible in the light curve.}
\end{center}
\end{figure*}

\begin{table*}
\caption{DNOs, lpDNOs and QPOs in V729 Sgr at the different states. The question mark represent uncertain case.}
 \begin{center}
 \large
   \begin{tabular}{ccccccc}\hline\hline
Date        &Outburst type   &$P_{DNO}$ (s)  &$P_{lpDNO}$ (s)     &$P_{QPO}$ (s)     &Filters \\\hline
2010 Jun 07 &rise/decline    &$-$	            &136?                &$-$                            &I\\
2011 Apr 04 &quiescence	     &$-$	            &$-$                 &378                             &N\\
2011 Apr 05 &quiescence	     &$-$	            &$-$                 &322                             &I\\
2014 Jul 20 &late rise	     &$-$	            &$-$                 &478                             &N\\
2014 Aug 05 &maximum	     &25.5   	        &49.9?               &308                             &N\\
2014 Aug 06 &early decline	 &36.2	            &$-$                 &231                             &N\\
2014 Aug 20 &late decline	 &$-$	            &$-$                 &587                             &N\\
2014 Aug 23 &quiescence	     &$-$	            &154                 &324                             &N\\
2015 Sep 09 &early rise?     &$-$	            &$-$                 &408                             &N\\
2016 Jul 06 &early decline?  &$-$	            &$-$                 &$-$                             &N\\\hline
\end{tabular}
\end{center}
\end{table*}

Our observations for V729 Sgr show distinct oscillations, and contain at least three outburst episodes (see Fig. 1). In order to explore the nature of these changes, a frequency analysis was performed by using the Fourier transform (FT) method described by Lomb (1976) and Scargle (1982). Due to the system has a low orbital inclination ($\sim 71^{\circ}$) (Cieslinski et al. 2000), the accretion disc is partly visible during the eclipses. Fig .1 show the simultaneous presence of both eclipses and the rapid brightness oscillations. To investigate more details of DNOs and QPOs, the data should be prepared by subtracting a mean eclipse profile trend from each light curve individually before computing the FT. Fig. 6 show the processed light curves near the outburst maximum (the left panel) and their Fourier spectrum diagrams (the right panel). Meanwhile, we have also calculated FTs of other light curves in order to compare them at the different state. Table 2 gives analysis results of DNOs and QPOs for each light curve. As mentioned above, only one outburst near the maximum was observed (see Table 2). Our analysis will first concentrate on this outburst. This system on 5 August 2014 was observed at the outburst maximum, and the dominant feature of the light curve is higher coherence and stronger short-period oscillation than the rise phase, and rapid oscillations are visible directly in the light curve. Moreover, its Fourier spectrum also shows larger amplitude and more peaks. The top-right panel of Fig. 6 displays the FT of the light curve at the outburst maximum. Note that the signal at $\sim 25.5$ s is only detected at the outburst maximum, implying that they may be true DNOs being associated with accretion events. The reason is that most of DNOs are generated from high accretion CVs such as DN in outburst and NLs. Returning to Table 2, there is another possible DNO at $\sim36.2$ s present at the early decline (see the lower panel of Fig. 6). The 36.2 and 25.5 s periodicities are close to $3:2$ ratio, which is similar to the harmonic of the synodic period in VW Hyi (Warner \& Woudt 2006; Woudt et al. 2010).

The possible oscillations at $\sim 136$ and $154$ s are classified as an lpDNO rather than a QPO because they are close to 4 times of the DNO period ($P_{lpDNOs}/P_{DNOs}\approx4$ for 36.2 s and $\approx5.7$ for 25.5 s (mean value)). Moreover, a signal at 49.9 s observed simultaneously with DNOs on 5 August 2014 was also identified as a possible lpDNO, but there is some ambiguity because the ratio $P_{lpDNOs}/P_{DNOs}$ is only $\sim1.96$.
The largest amplitude modulations in all FTs are not caused by a normal QPO but related to the sine-fitting curve in Fig.6 and Fig. 7. However, it is not clear whether the sinusoidal signal in the light curves is true or not. A typical example with an obvious QPO is given in Fig. 7, which shows a 408 s QPO. For this individual case, $P_{QPO}/P_{DNO}\approx15.94$ being very close to the typical value 16.

\section{Conclusion}
We have presented the photometric results of an
eclipsing dwarf nova V729 Sgr using our observations together with AAVSO data.
Our analysis is focus on the outburst properties and rapid oscillations in brightness.
The main conclusions are summarized as follows:

(i)  Based on our new observations and the AAVSO data of $\sim15$ yr we have found that the outburst amplitude is in the range of $0.6-2.2$ mag. The power spectrum of AAVSO data indicates that there are some prominent peaks corresponding to the typical recurrence times, which are $\sim26.5$ d for normal outburst and $9.1-15.6$ d for short outburst. By comparing the observational data of V729 Sgr and the K-P relations, we believe that the K-P relation represents general properties of DN outbursts and is model dependent.
Moreover, the long-term outburst curves display the presence of the small-amplitude outbursts and dips, which is very similar to the "stunted" outbursts found in some NLs. Further examination revealed that V729 Sgr may be undergoing Type A outburst. The rise time for Type A was estimated as $\sim1.76$ d, in consistent with observations roughly.
As a Z Cam type star, V729 Sgr exhibits the diverse outburst behaviors containing the normal outbursts in DN and the "stunted" outbursts in NLs. Combining the distribution diagram of outburst amplitudes in CVs we suggest that it is an intermediate between DN and NLs. We also constrain the mass transfer rate in this system to $1.59\times10^{-9}<\dot{M}_{2}<5.8\times10^{-9}M_{\odot}yr^{-1}$ by considering the critical mass transfer rate and the outburst properties in V729 Sgr.

(ii)  We have detected the rapid oscillations in V729 Sgr, the first to be reported for this star. A DNO at $\sim 25.5$ s in V729 Sgr was found during outburst, indicating that it is related to the accretion from the inner regions of the accretion disc on to the surface of white dwarf. The classification of the oscillations at $\sim 136$ and $154$ s as lpDNOs is based on the relation $P_{lpDNOs}/P_{DNOs}\approx4$. For QPOs, a typical example is that there is a clear QPO with a period of 408 s present in the 2015 September 9 light curve, and $P_{QPO}/P_{DNO}\approx15.69$. To ascertain these conclusions, further observations should be encouraged.

\bigskip

\vskip 0.3in \noindent
This work is supported by the Chinese Natural Science Foundation (Grant No. 11325315, 11611530685, 11573063 and 11133007), the Strategic Priority Research Program ''The Emergence of Cosmological Structure'' of the Chinese Academy of Sciences (Grant No. XDB09010202) and the Science Foundation of Yunnan Province (Grant No. 2012HC011). This study is supported by the Russian Foundation for Basic Research (project No. 17-52-53200).
New CCD photometric observations of V729 Sgr were obtained with the 2.15-m Jorge Sahade telescope (JST) in Complejo Astron\'omico E1 Leoncito (CASLEO), San Juan, Argentina. We gratefully acknowledge the numerous observers contributing the data to the AAVSO Database. Finally, we thank the anonymous referee for those helpful comments and suggestions.



\begin{thebibliography}{}

\bibitem[Andronov et al. (2008)]{2008....28P} Andronov, I. L., Chinarova, L. L., Han, W., Kim, Y., \& Yoon, J.-N., \ 2008, \aap, 486, 855
\bibitem[Buat (2001)]{2001....28P} Buat-M{\'e}nard, V., Hameury, J.-M., \& Lasota, J.-P. \ 2001, \aap, 366, 612
\bibitem[Cannizzo (1993)]{1993....28P} Cannizzo, J. K. \ 1993, \apj, 419, 318
\bibitem[Cieslinski et al. (2000)]{2000PASP....112..349C} Cieslinski, D., Steiner, J. E., Pereira, P. C. R., \& Pereira, M. G. \ 2000, PASP, 112, 349
\bibitem[Cieslinski et al. (1997)]{1997AAS....124..55C} Cieslinski, D., Jablonski, F. J., \& Steiner, J. E. \ 1997, A\&AS, 124, 55
\bibitem[Cieslinski et al. (1998)]{1998AAS....131..119C} Cieslinski, D., Steiner, J. E., \& Jablonski, F. J. \ 1998, A\&AS, 131, 119
\bibitem[Eggleton (1983)]{1983....28P} Eggleton, P. P. \ 1983, \apj, 268, 368
\bibitem[Ferwerda (1934)]{1934BAIN....7..166F} Ferwerda, J. G. \ 1934, Bull. Astron. Inst. Netherlands, 7, 166
\bibitem[Frank (2002)]{2002....128F} Frank J., King A., Raine D., \ 2002, Accretion Power In Astrophysics, Cambridge University Press, 128
\bibitem[Gies (2013)]{2013....64G} Gies D. R. et al. \ 2013, \apj, 775, 64
\bibitem[Honeycutt (1995)]{1995....28P} Honeycutt R. K., Robertson J. W., Turner G. W., \ 1995, \apj, 446, 838
\bibitem[Honeycutt (1998)]{1998....28P} Honeycutt R. K., Robertson J. W., Turner G. W., \ 1998, \aj, 115, 2527
\bibitem[Honeycutt (2001)]{2001....473P} Honeycutt R. K., \ 2001, PASP, 113, 473
\bibitem[Honeycutt (2014)]{2014....10P} Honeycutt R. K., Kafka S., Robertson J. W., \ 2014, \apj, 147, 10
\bibitem[Howell (2000)]{2000....28P} Howell S. B., Ciardi D. R., Dhillon V. S., SkidmoreW., \ 2000, \apj, 530, 904
\bibitem[Kafka \& Honeycutt (2005)]{2005....28P} Kafka S., Honeycutt R. K., \ 2005, \aj, 130, 742
\bibitem[Kotko \& Lasota (2012)]{2012....28K} Kotko, I \& Lasota, J.P., \ 2012, \aap, 545, 115
\bibitem[Kukarkin \& Parenago (1934)]{1934....28P} Kukarkin, B. W., \& Parenago, P. P., \ 1934, Var. Star. Bull., 4, 44
\bibitem[Lasota (2001)]{2001....28P} Lasota, J.-P. \ 2001, NewA Rev., 45, 449
\bibitem[Lomb (1976)]{1976....28P} Lomb, N. R., \ 1976, Ap\&SS, 39, 447L
\bibitem[Osaki (1996)]{1996PASP....108..39O} Osaki, Y. \ 1996, PASP Rev, 108, 39
\bibitem[Pretorius et al. (2006)]{2006MNRAS....368..361P} Pretorius M. L., Warner B., Woudt P. A., \ 2006, \mnras, 368, 361
\bibitem[Ramsay (2016)]{2016....28P} Ramsay G., Hakala P., Wood M. A., Howell s. b. et al., \ 2016, \mnras, 455, 2772
\bibitem[Richter \& Braeuer (1989)]{1989....28P} Richter, G. A. \& Braeuer, H.-J., \ 1989, AN, 310, 413
\bibitem[Scargle (1982)]{1982....28P} Scargle, J. D., \ 1982, \apj, 263, 835
\bibitem[Smak (1984)]{1984....28P} Smak, J. \ 1984, Acta Astron., 34, 161
\bibitem[Smak (2001)]{2001....28P} Smak, J. \ 2001, Acta Astron., 51, 279
\bibitem[Szkody (2013)]{2013....28P} Szkody, P., Albright, M. et al. \ 2013, PASP, 125, 1421
\bibitem[van Gent (1932)]{1932....28P} van Gent, H. \ 1932, BAN, 6, 163
\bibitem[van Paradijs (1985)]{1985....28P} van Paradijs, J. \ 1985, \aap, 144, 199
\bibitem[Warner (2004)]{2004PASP....116..115W} Warner B., \ 2004, PASP, 116, 115
\bibitem[Warner (1995)]{1995....28P} Warner, B.\ 1995, Cataclysmic Variable Stars Cambridge Astrophysics Series (Cambridge: Cambridge Univ. Press), 28
\bibitem[Warner \& Woudt (2002)]{2002MNRAS....335..84W} Warner B., Woudt P. A., \ 2002, \mnras, 335, 84
\bibitem[Warner et al. (2003)]{2003MNRAS....334..1193W} Warner B., Woudt P. A., Pretorius M. L., \ 2003, \mnras, 334, 1193
\bibitem[Warner \& Woudt (2006)]{2006MNRAS....367..1562W} Warner B., Woudt P. A., \ 2006, \mnras, 367, 1562
\bibitem[Warner \& Pretorius (2009)]{2009MNRAS....383..1469W} Warner B., Pretorius M. L., \ 2008, \mnras, 383, 1469
\bibitem[Woudt \& Warner (2002)]{2002MNRAS....333..411W} Woudt P. A., Warner B., \ 2002, \mnras, 333, 411
\bibitem[Woudt \& Warner (2009)]{2009MNRAS....400..835W} Woudt P. A., Warner B., \ 2009, \mnras, 400, 835
\bibitem[Woudt \& Warner (2010)]{2010MNRAS....401..500W} Woudt P. A., Warner B. et al., \ 2010, \mnras, 401, 500
\end{thebibliography}
\end{document}